\documentclass{article}

\PassOptionsToPackage{numbers, compress}{natbib}
\usepackage[final]{neurips_2024}

\usepackage[utf8]{inputenc} 
\usepackage[T1]{fontenc}    
\usepackage{hyperref}       
\usepackage{url}            
\usepackage{booktabs}       
\usepackage{amsfonts}       
\usepackage{nicefrac}       
\usepackage{microtype}      
\usepackage{xcolor}         
\usepackage[pdftex]{graphicx}
\usepackage{subcaption}
\title{Demo: Interactive Visualization of Semantic Relationships in a Biomedical Project's Talent Knowledge Graph}

\author{
  Jiawei Xu \\
  School of Information\\
  University of Texas at Austin\\
  \texttt{jiaweixu@utexas.edu} \\
  \And
  Zhandos Sembay \\
  School of Medicine \\
  University of Alabama at Birmingham\\
  \texttt{zsembay8@uab.edu} \\
  \And
  Swathi Thaker \\
  School of Medicine \\
  University of Alabama at Birmingham\\
  \texttt{snthaker@uab.edu} \\
  \And
  Pamela Payne-Foster\\
  School of Medicine \\
  University of Alabama \\
  \texttt{pfoster@ua.edu} \\
  \And
  Jake Yue Chen \\
  School of Medicine \\
  University of Alabama at Birmingham \\
  \texttt{jakechen@uab.edu} \\
  \And
  Ying Ding \\
  School of Information \\
  University of Texas at Austin \\
  \texttt{ying.ding@ischool.utexas.edu}
}

\begin{document}

\maketitle

\begin{abstract}
We present an interactive visualization of the Cell Map for AI Talent Knowledge Graph (CM4AI TKG), a detailed semantic space comprising approximately 28,000 experts and 1,000 datasets focused on the biomedical field. Our tool leverages transformer-based embeddings, WebGL visualization techniques, and generative AI, specifically Large Language Models (LLMs), to provide a responsive and user-friendly interface. This visualization supports the exploration of around 29,000 nodes, assisting users in identifying potential collaborators and dataset users within the health and biomedical research fields. Our solution transcends the limitations of conventional graph visualization tools like Gephi, particularly in handling large-scale interactive graphs. We utilize GPT-4o to furnish detailed justifications for recommended collaborators and dataset users, promoting informed decision-making. Key functionalities include responsive search and exploration, as well as GenAI-driven recommendations, all contributing to a nuanced representation of the convergence between biomedical and AI research landscapes. In addition to benefiting the Bridge2AI and CM4AI communities, this adaptable visualization framework can be extended to other biomedical knowledge graphs, fostering advancements in medical AI and healthcare innovation through improved user interaction and data exploration. The demonstration is available at: \href{https://jiawei-alpha.vercel.app/}{https://jiawei-alpha.vercel.app/}.
\end{abstract}

\section{Introduction}

This paper presents an interactive visualization of the Bridge2AI - Cell Maps for AI Data Generation Project Talent Knowledge Graph (CM4AI TKG)~\cite{clark2024cell,xu2023cross} semantic space. The Bridge2AI project \cite{suran2022new} assembles professionals from biomedicine, computer science, and social sciences to develop and curate high-quality biomedical datasets critical for AI-driven research and transformative healthcare advancements. The CM4AI TKG is designed to assist in identifying suitable potential users for these datasets~\cite{rigden20232023} and locating relevant collaborators within the extensive biomedical domain. Our visualization tool supports the exploration of the distribution of CM4AI researchers and datasets within the broader biomedical and genomics landscape, thereby aiding the discovery of future collaborators and users. We use Large Language Models (LLM) to offer suggestions for potential collaborators and users for each researcher and biomedical dataset in the CM4AI TKG.

Our visualization application is highly adaptable, incorporating node representations via transformer-based semantic embeddings \cite{Singh2022SciRepEvalAM}, dimensionality reduction techniques like t-SNE and UMAP \cite{Maaten2008VisualizingDU, McInnes2018UMAPUM}, and large-scale responsive node visualization through PixiJS~\cite{van2015learn}. This configuration allows for the future integration of additional functionalities. Compared to widely used graph visualization tools, such as Gephi \cite{bastian2009gephi} and Cytoscape \cite{shannon2003cytoscape}, PixiJS efficiently handles and visualizes large datasets while offering excellent cross-platform capabilities. Furthermore, we incorporate LLM to enhance user understanding and provide explainability. This interactive visualization framework can be adapted for other medical domain knowledge graphs, offering an interactive interface that improves user accessibility.

The CM4AI TKG was extracted from the PubMed Knowledge Graph \cite{xu_building_2020} and Semantic Scholar \cite{kinney2023semantic}. It includes 2 million papers, 44,000 authors, 1,179 biomedical datasets~\cite{rigden20232023} used in these studies, and bio-entities mentioned in the papers. The interactive visualization of the CM4AI TKG provides an intuitive, user-friendly interface for exploring the knowledge graph. Utilizing semantic information from paper titles and abstracts \cite{Singh2022SciRepEvalAM}, our system maps a two-dimensional semantic space, indicating the relative positions of researchers and datasets within the CM4AI knowledge domain. Distances between researchers and datasets reflect their similarities. Users can search for datasets or talents, navigate the space with zoom functionality, and access detailed author and dataset information.

The following sections cover the data and methodology used in constructing the CM4AI TKG, along with an outline of its functionalities and practical applications.

\section{Data and Methods}

\subsection{Data Preparation}

The CM4AI TKG compiles data related to 121 core researchers participating in the Bridge2AI project, including 35 researchers identified as core members of CM4AI. This dataset encompasses these researchers' 44,000 co-authors and approximately 2.05 million publications in total. Out of these, we specifically employed 10,011 publications associated directly with the core researchers. 

To construct the dataset, we first collected ORCID~\cite{haak2012orcid} identifiers for the 121 core researchers and manually confirmed their identities using Semantic Scholar to ensure accuracy. Following this, we matched these researchers within the PubMed Knowledge Graph \cite{xu_building_2020} to gather comprehensive data on their publications, co-authors, datasets used, and mentioned biological entities. To enhance relevance and focus, we excluded any authors without publications after 2020, refining our dataset to comprise 28,000 active researchers, referred to as ‘talents,’ for visualization purposes.

\subsection{Author and Dataset Representation}

To represent authors and datasets, we employed Specter2, a state-of-the-art BERT-based encoder~\cite{Singh2022SciRepEvalAM, kenton2019bert}, to transform the titles and abstracts of all 2.05 million papers into 768-dimensional embedding vectors. For author representation, we aggregated these embeddings by weighing the author's position in each publication. Specifically, the first and last authors received a weight of 1, while a \( k \)-th author was assigned a weight of \( \frac{1}{k} \). Authors beyond the 10th position were assigned a uniform weight of \( \frac{1}{10} \). Dataset representations were obtained by aggregating embeddings from papers that utilized the respective datasets. To identify potential collaborators, we selected the top 30 researchers for each author—individuals with whom they had never collaborated—by computing cosine similarity of the embeddings. Similarly, for each dataset, we identified the top 150 researchers who had not used the dataset as potential users.

We utilized GPT-4o~\cite{achiam2023gpt} to provide justifications for these recommendations. For collaborator recommendations, we inputted to the model five recent and five most-cited papers since 2017 for each author, along with metadata such as title, journal, citation count, and publication year~\cite{hutchins2019nih}. The model then generated justifications highlighting the potential benefits of collaboration. For recommending dataset users, we provided the model with an author's recent and highly cited papers and accompanying dataset descriptions, prompting the model to justify why these users should consider the dataset.

\subsection{Visualization}

For visualizing the data, we utilized PixiJS~\cite{van2015learn}, a versatile 2D WebGL renderer, capable of efficiently displaying large numbers of nodes in a two-dimensional space. Unlike traditional graph visualization tools such as Gephi \cite{bastian2009gephi} and Cytoscape \cite{shannon2003cytoscape}, PixiJS takes advantage of WebGL technology to handle and render large datasets, ensuring high cross-platform compatibility. This allows users to explore the visualization effortlessly through any modern web browser. The web application was developed using TypeScript and Svelte, with visualization and hosting code adapted from an open-source anime recommendation project, Sprout \cite{primozic_ameobeasprout_2024}. Additionally, an Oracle APEX visualization is available, providing detailed information such as each talent's publication history, accessible at: \href{https://cm4ai.org/ckg}{https://cm4ai.org/ckg}.

To ensure a well-organized layout, crucial for effective embedding visualization and user experience, we experimented with various dimensionality reduction techniques and layout configurations using Emblaze \cite{cmu_data_interaction_group_cmudigemblaze_2024}. This tool facilitates the visual comparison of embedding spaces and offers built-in dimensionality reduction methods like t-SNE and UMAP \cite{Maaten2008VisualizingDU, McInnes2018UMAPUM}. These methods were used to condense the 768-dimensional embeddings of authors and datasets into two-dimensional coordinates. We tuned the parameters of t-SNE and UMAP to achieve an optimal layout that enhances visualization quality and user interaction.

\section{Function and Use Cases}

\begin{figure}[ht]
  \centering
  \begin{subfigure}[t]{.80\textwidth}
    \centering
    \includegraphics[width=\textwidth]{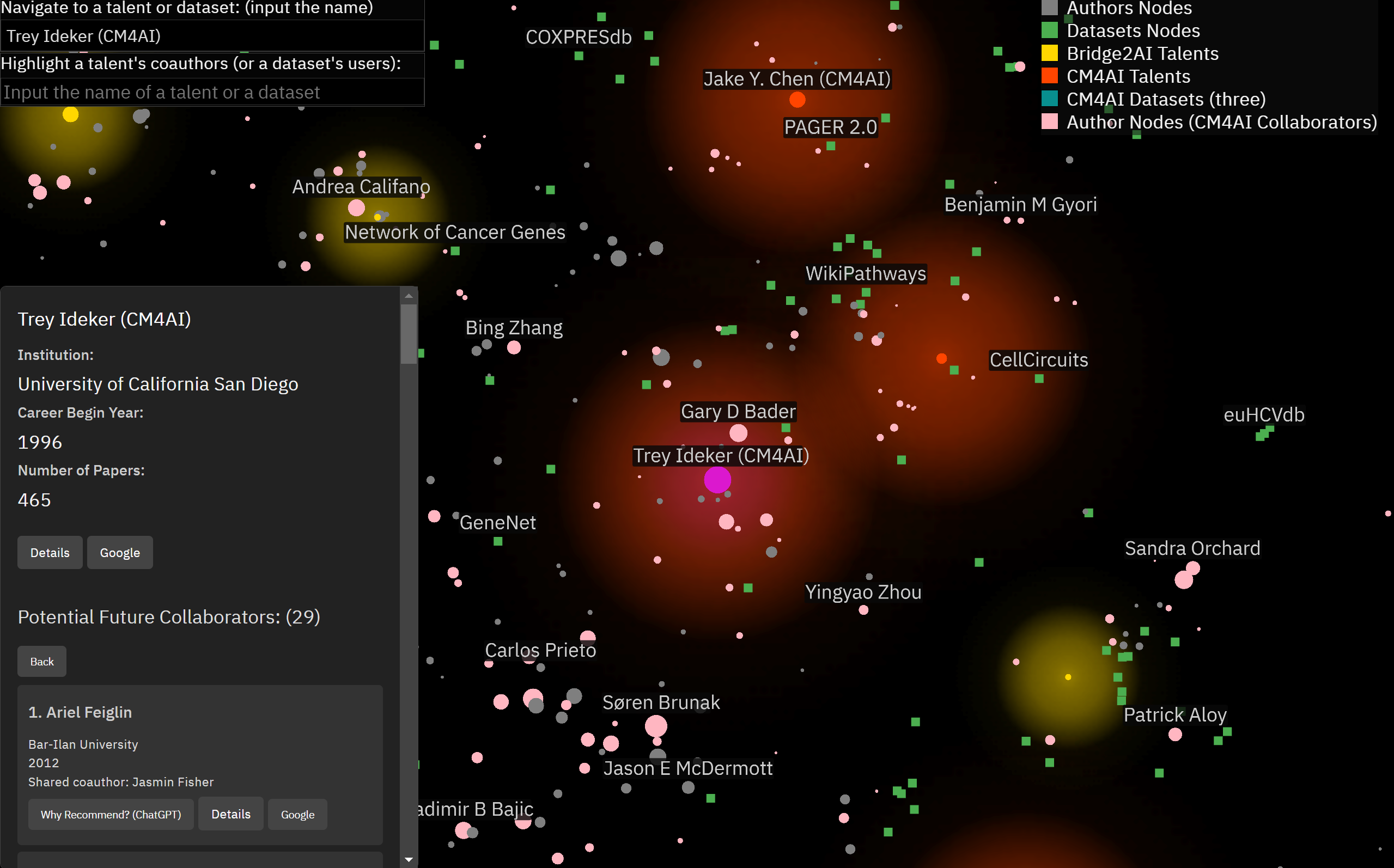}
    \caption{Information Window for a Talent (Trey Ideker)}
    \label{fig:figure1a}
  \end{subfigure}%
  \hspace{0.05\textwidth} 
  \begin{subfigure}[t]{.80\textwidth}
    \centering
    \includegraphics[width=\textwidth]{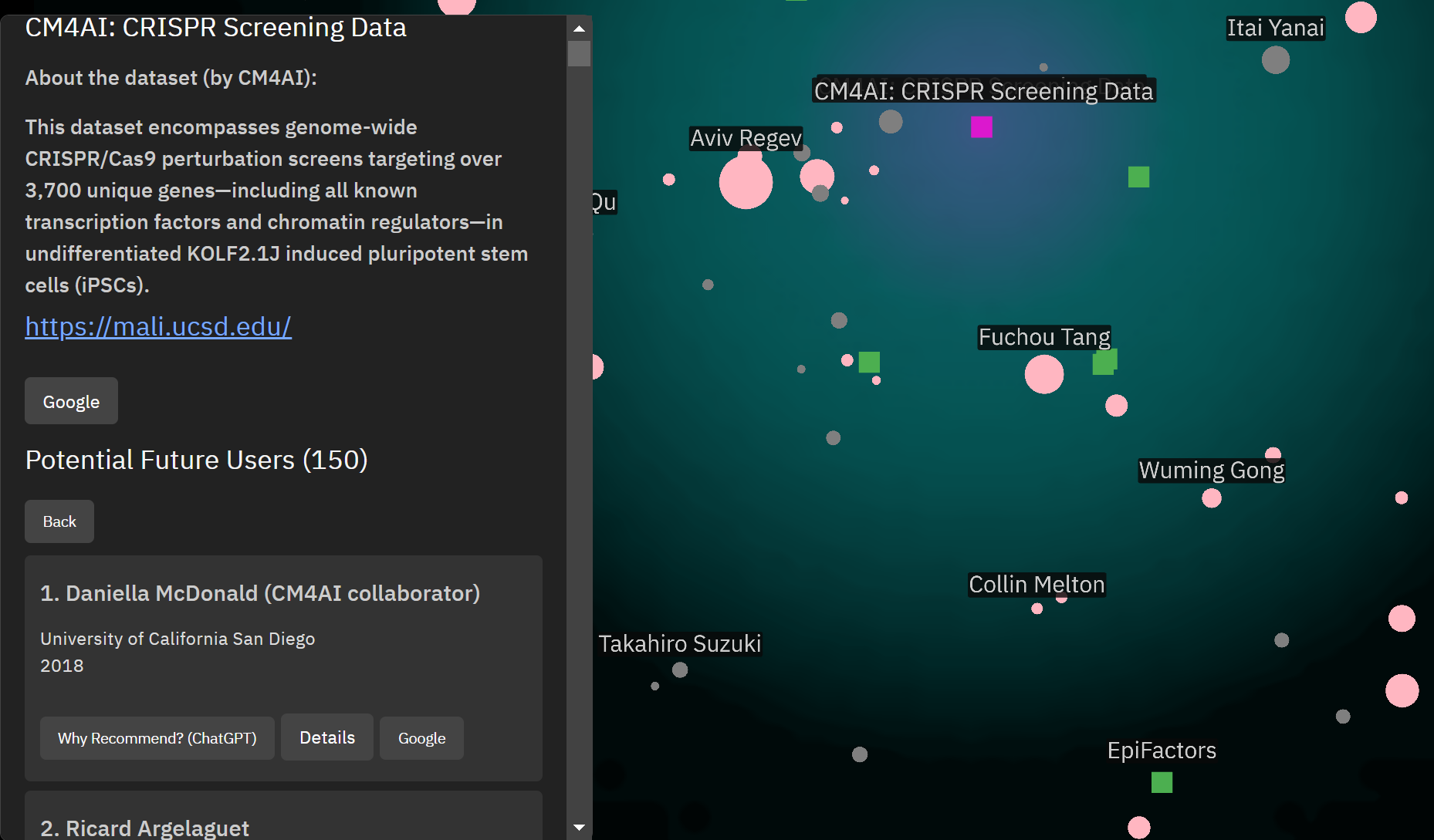}
    \caption{Information Window for a Dataset (CRISPR Screening Data)}
    \label{fig:figure1b}
  \end{subfigure}
  \caption{Information windows for different items: (a) a talent, (b) a dataset.}
  \label{fig:figure1}
\end{figure}

The CM4AI TKG semantic space visualizes talents and datasets as nodes within a two-dimensional space, as illustrated in Figure~\ref{fig:figure1}. Node size reflects the number of publications linked to each talent (log scale), while node shapes and colors distinguish between node types: dataset nodes are depicted as squares and talent nodes as circles. This interactive interface allows users to drag, scroll, zoom, hover over, and search nodes across approximately 28,000 talent nodes and 1,179 dataset nodes. 
The system provides two main functionalities: (1) Exploration of existing positions in the knowledge space and (2) Recommendation of potential collaborators or dataset users with justifications provided by GPT-4o.

\begin{figure}[ht]
  \centering
  \begin{subfigure}[t]{0.85\textwidth}
    \centering
    \includegraphics[width=\textwidth]{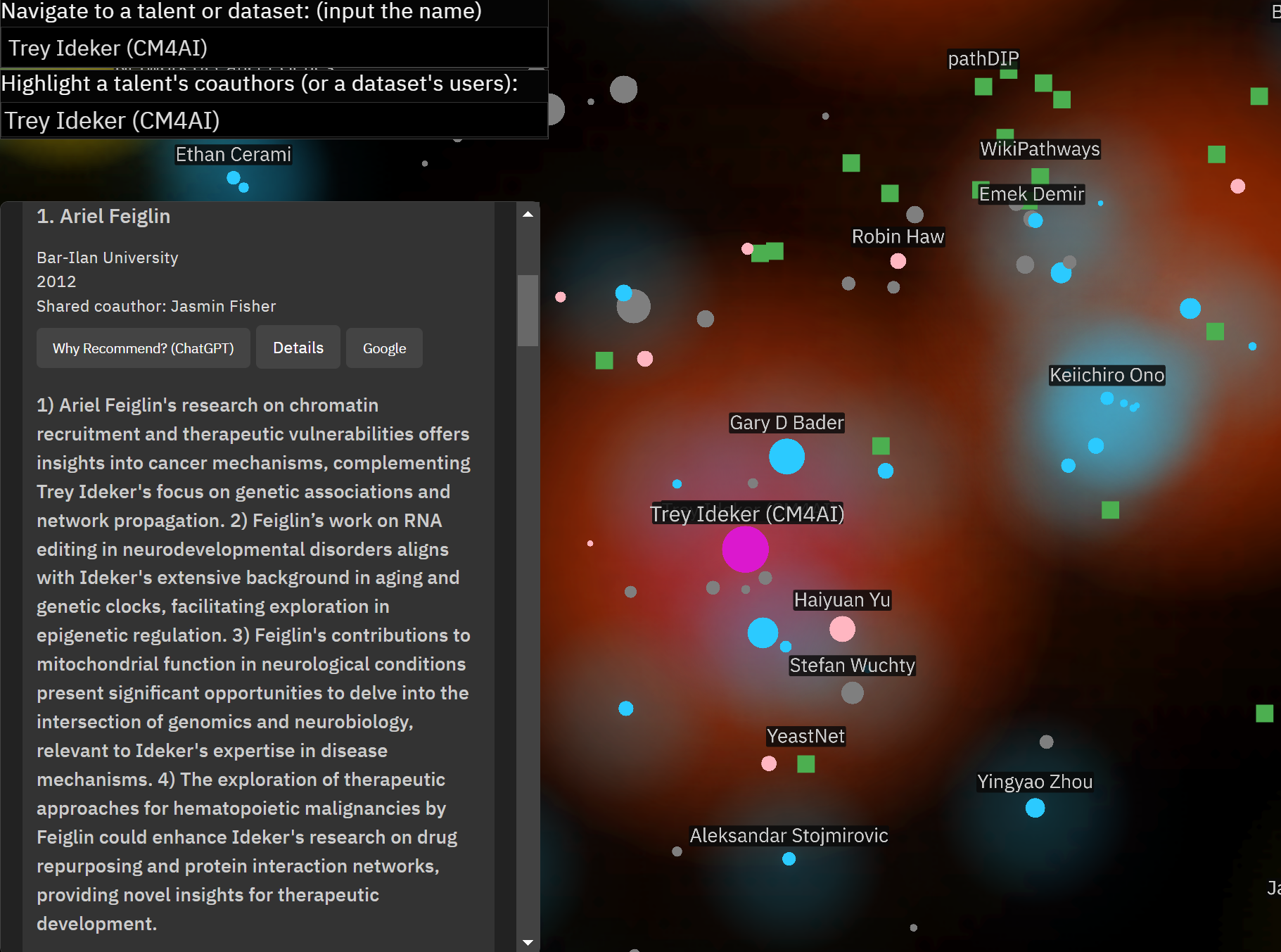}
    \caption{LLM's Justification for Recommending Ariel Feiglin to Trey Ideker}
    \label{fig:figure2a}
  \end{subfigure}%
  \hspace{0.05\textwidth} 
  \begin{subfigure}[t]{0.85\textwidth}
    \centering
    \includegraphics[width=\textwidth]{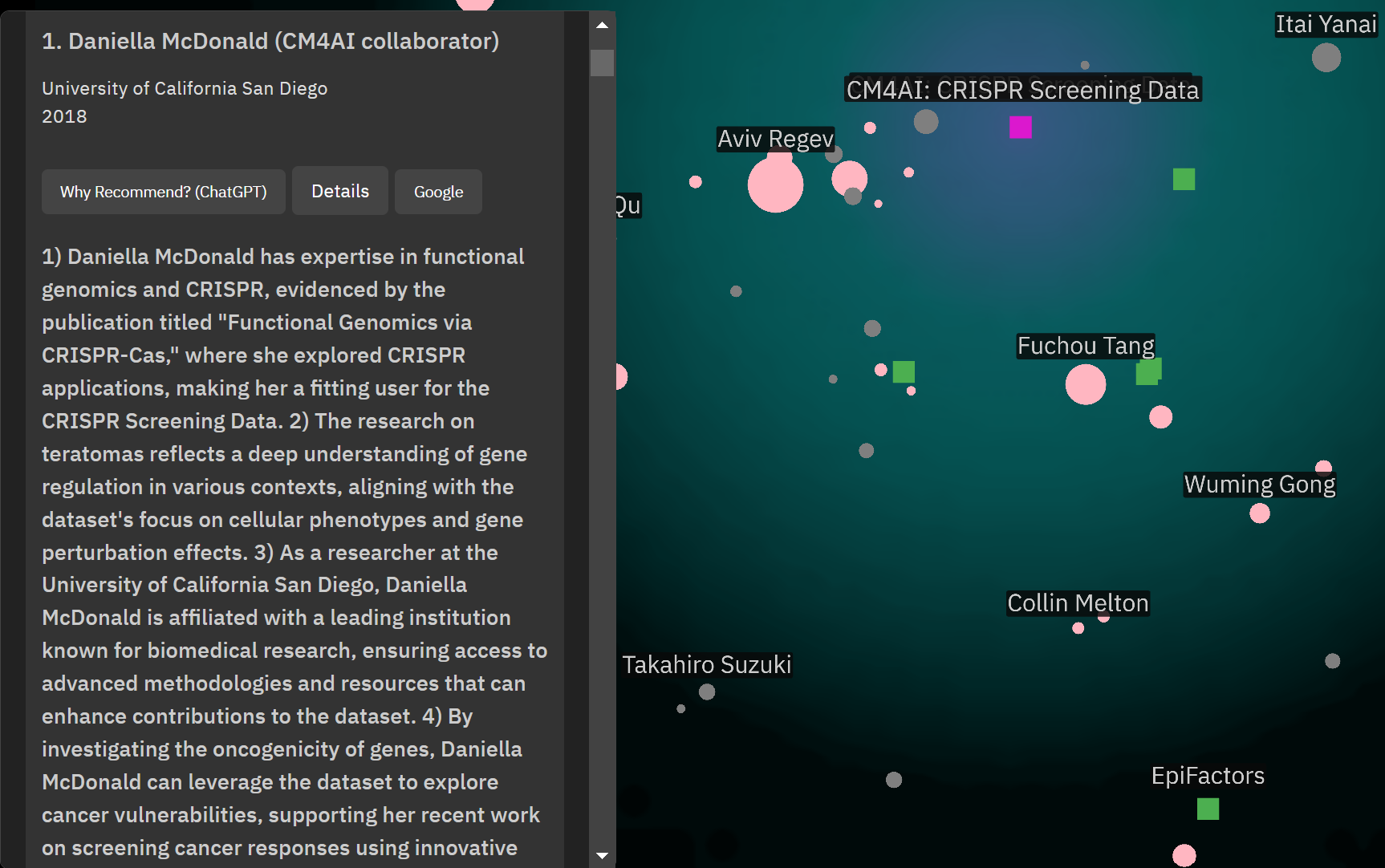}
    \caption{LLM's Justification for Recommending Daniella MacDonald to CRISPR Screening Data}
    \label{fig:figure2b}
  \end{subfigure}
  \caption{LLM's Justifications for Recommendations: (a) For Trey Ideker, (b) For CRISPR Screening Data}
  \label{fig:figure2}
\end{figure}

\textbf{Explore Existing Position in the Knowledge Space}. Users can efficiently locate and explore a specific dataset or talent by entering its name in the search box. The system then provides a list of candidates; once selected, such as ‘Trey Ideker’ shown in Figure~\ref{fig:figure1a}, the view zooms to this talent, presenting an information window with key details such as institution, number of publications, and career start year. By clicking the ‘Detail’ button, users are directed to an Oracle APEX interface to review detailed publication histories. Similarly, searching for a dataset reveals detailed information in a popup window, as demonstrated with ‘CRISPR Screening Data’ in Figure~\ref{fig:figure1b}. The ‘Explore Existing Collaborators or Users’ feature helps users understand current collaboration and usage patterns. By entering a talent’s name, such as ‘Trey Ideker’, in the second search box, users can highlight the nodes of talents who have previously collaborated with that individual, creating a visually distinct starry effect (Figure~\ref{fig:figure2a}). This facilitates comparisons between historical and potential future collaborations.

\textbf{Use of Generative AI for Informed Recommendations}. The system also provides lists of recommended collaborators or dataset users, accompanied by LLM-generated justifications explaining why these individuals are suggested. By clicking the ‘Why Recommend?’ button, users can view these justifications. Users can explore these potential collaborators by clicking their names, redirecting the view to their respective nodes. This feature leverages the reasoning capabilities of LLMs to enhance the usefulness of recommendations. In Figure~\ref{fig:figure2a}, for instance, the model justifies recommending Ariel Feiglin to Trey Ideker, and in Figure~\ref{fig:figure2b}, it provides reasons for suggesting Daniella MacDonald as a user for the CRISPR Screening Data.
\section{Summary}
The current demonstration offers a rapid-response interface for users to interact with the CM4AI Talent Knowledge Graph (TKG). It integrates user recommendations with explanations generated by LLMs that are based on detailed background information. Additionally, it displays profiles of experts and datasets with further details via both an information window and a database visualization. The semantic space of the CM4AI TKG provides an intuitive and user-friendly interface, which showcases the growth of the Bridge2AI and CM4AI community within the expansive biomedical landscape. This WebGL-based visualization technology is not only advantageous for the Bridge2AI and CM4AI communities but can also be adapted for other domain-specific knowledge graphs containing hundreds of thousands of nodes. There are some limitations to be addressed in this work: (1) Author name disambiguation and normalization~\cite{torvik2009author}. The user experience can be affected by inconsistencies in author representation. Some authors may not be accurately disambiguated, leading to discrepancies, such as variations in name representation, where some authors are reported with full names while others only have initials. (2) Subjective evaluation of visualization. The evaluation of the final visualization is relatively subjective. Using dimensionality reduction methods like t-SNE or UMAP to determine node coordinates, the current assessment is based on the visual clustering and interpretability of node positions. Future work should focus on developing objective methods to evaluate the visualization's effectiveness.

Overall, this visualization pipeline provides considerable advantages for medical applications that require visualizing extensive datasets and customizing frameworks to incorporate LLM reasoning. By continually refining these capabilities, we can enhance the utility of large-scale knowledge graphs or vector (embedding) data. This advancement drives progress in medical AI and research by offering user-friendly information representation, thereby facilitating better data interpretation and decision-making in healthcare innovations.

\section*{Acknowledgments}
We would like to acknowledge the following funding supports: NIH 1OT2OD032742-01, NIH OTA-21-008, NIH OT2OD032581, NSF 2333703, NSF 2303038.

\bibliography{citation}

\end{document}